\title{Epistemic relativity:\\
An experimental approach to physics}
\author{Bartolom\'e Coll\footnote{Email: bartolome.coll@uv.es} \\
Relativistic Positioning Systems\\
Department of Astronomy \& Astrophysics\\
University of Valencia\\
c/ Dr. Moliner 50, 46100 Burjassot (Valencia), Spain}
\date{\today}
\documentclass[11pt,english]{article}
\usepackage[]{graphicx}
\usepackage{url}
\RequirePackage{fix-cm}
\begin{document}
\maketitle
\begin{abstract}
The recent concept of {\em relativistic positioning system} (RPS) has opened the possibility of making Relativity the {\em general standard frame} in which to state any physical problem, theoretical or experimental. 

Because the velocity of propagation of the information is finite, {\em epistemic relativity} proposes to integrate the 
physicist as a real component of every physical problem, taking into account explicitly {\em what} information, {\em when} and {\em where}, the physicist is able to know. This leads naturally to the concept of {\em relativistic stereometric system} (RSS), allowing to measure the intrinsic properties of physical systems. Together, RPSs and RSSs complete the notion of {\em laboratory} in general relativity, allowing to perform experiments in finite regions of any space-time.

Epistemic relativity incites the development of relativity in new open directions: advanced studies in RPSs and RSSs, intrinsic characterization of gravitational fields, composition laws for them, construction of a finite-differential geometry adapted to RPSs and RSSs, covariant approximation methods, etc. Some of these directions are sketched here, and some open problems are posed. 
\end{abstract}
\noindent I want the organizers of this seminar, Dirk P\"utzfeld and Claus L\"ammerzahl, to know how much I appreciated their inviting me to talk about my ideas on this subject. It is also a pleasure to thanks the Wilhelm und Else Heraeus-Stiftung for their kind hospitality.
%
%
\section{Introduction}
As everyone knows, General Relativity is a theory of the gravitational field but, also and primarily, a theory of the space-time. During the last century, it has been largely proved experimentally that the gravitational field, as well as the space-time, is better described by it than by Newtonian theory.

Nevertheless, by natural reasons, the environments in which most of the experiments have taken place have been either static (e.g. Earth surface), or stationary (e.g. aircrafts or satellites in circular orbits), or periodic (e.g. satellites in elliptic orbits), or quasi-periodic with almost-periodic or constant non-periodic parts (e.g. Solar system ephemeris or radiation by black hole accretion disks). These environments allow trivializing some important specificities of the relativistic space-time, especially some related to the finite velocity of propagation of the information. 

Thus, for example, in these particular environments, the deterministic character of Einstein equations seems to be {\em predictive}, like the (Laplace) one of Newtonian theory. But out of these environments, because Einstein equations can be associated to a hyperbolic system%
\footnote{The differential operator associated to the $n(n+1)/2$ Einstein equations is not hyperbolic, but degenerate. Nevertheless, because $n$ of these equations (constraint equations) are involutive with respect to the other $n(n-1)/2$ ones (evolution equations), one can supplement this last system with $n$ suitable additional equations (coordinate conditions). It is the differential operator of this new system of $n(n+1)/2$ equations for the metric coefficients which may be made hyperbolic with a suitable choice of coordinate conditions (for example, with harmonic ones).} %
with influence domain%
\footnote{The influence domain for an initial instant (local spatial hypersurface) is the domain where the solution exists and is unique for every initial (Cauchy) data on this initial instant.} %
attached to the velocity of light, the initial data on an initial instant cannot be {\em physically} known but at or after the cusp of the influence domain, so that Einstein determinism is generically {\em retrodictive},%
\footnote{Retrodictive, as opposed to predictive, means here that one cannot but {\em verify afterwards} that the physical quantities measured {\em in} the influence domain agree with the initial data received {\em at or after} the cusp of this domain.} %
not predictive. Nevertheless, most of the problems related to initial conditions in relativity, even for generic environments, have been thought as predictive ones, as if someone at the initial instant were able to know the initial data. This manner of thinking, strongly attached to the evolutive Newtonian point of view%
\footnote{In Newtonian physics, the contents of the three-dimensional space at every instant is supposed known or knowable at that instant, so that the evolutive equations describe the dependence in time of this three-dimensional contents.} %
but physically inadmissible in relativistic generic environments, masks the concepts that we should develop in order to be able to {\em describe physically generic environments in relativity}.

The aim of {\em epistemic relativity} is to give us such ability.%
\footnote{Epistemic relativity was first presented at the GraviMAS FEST workshop, in honor of Llu\'{\i}s Mas, Mallorca, Spain, 2008, \url{http://www.uib.es/depart/dfs/GRG/GraviMAS_FEST/}. See also \cite{Coll:2013}.} %
Section 2 is devoted to specify its basic ingredients. 
The first of these ingredients is constituted by {\em relativistic positioning systems},%
\footnote{For the genesis of the concept of relativistic positioning systems, see for example \cite{Coll:2013}.} %
of which {\em auto-locating systems} and {\em autonomous positioning systems} are the more interesting ones. They are the object of Section 3.
The second basic ingredient is constituted by {\em relativistic stereometric systems},%
\footnote{They were relativistic stereometric systems which, joined with relativistic positioning systems, suggested the idea of epistemic relativity. This is why relativistic stereometry and epistemic relativity were conceived conjointly.} %
which allow to detect intrinsic properties of physical systems. Their notion and first physical results are presented in Section 4. By the way, epistemic relativity plays a heuristic role giving new meanings to already known subjects or opening new ones. Section 5 shows how the {\em intrinsic characterization} of metrics may help epistemic relativity to identify gravitational fields and, finally, Section 6 introduces a {\em finite-differential geometry} in which distance function and metric ought to play symmetric roles.

I believe that in epistemic theory, as an incipient theory, what is interesting are the basic concepts. This is why I have restricted the bibliography to them.%
\footnote{Some subjects related to epistemic relativity have been the object of interesting developments (e.g. positioning systems, relativistic geometric optics, intrinsic characterization of non-vacuum metrics, Regge calculus) not directly related to the basis of this theory. For this reason, the corresponding bibliography is absent here.}
%
%
\section{Epistemic relativity}
Being the best theory of the space-time in which all the physical phenomena take place, relativity ought to be able to describe any physical experiment in terms of its proper, relativistic, concepts, regardless of the quantitative evaluation of the experiment for which, in many cases, Newtonian calculations could suffice.

Also, being the best theory of the gravitational field, relativity ought to propose experiences and methods of measurement of general 
gravitational fields.

But today, such descriptions or proposals are conspicuous by their absence. In fact, we do not know how to do them. Thus it seems evident that
{\em relativity needs to develop a proper experimental approach to the physical world.} 
%
%
\subsection{The notion of epistemic relativity}
Fortunately, we already know the conceptual basic ingredients for such a development. Before to describe them, the idea of a `relativistic experimental approach' needs to be more explicit.

In relativity, a large number of scientific works analyze physical and geometrical properties of the space-time, but
\begin{itemize}
\item[$\bullet$] do not integrate the physicist as a part of them, and
\item[$\bullet$] forget implicitly that
\vspace{-1.mm}
\begin{itemize}
\vspace{-1.mm}
\item[*] information is energy,
\vspace{-1.mm}
\item[*] neither the density of energy nor its velocity of propagation can be infinite in relativity.
\vspace{-2.5mm}
\vspace{-1mm}
\end{itemize}
\end{itemize}
\vspace{1.5mm}
Many of these properties of the space-time may be analyzed by a geometer on his desk, but to be known by an experimental physicist,%
\footnote{In all this text, the worlds `physicist', `observer' or `user' denote any person or device able to receive the pertinent information, to record and to analyze it and to perform the actions and computations needed for the problem in question. For short, we shall refer to any of them as `it'.} %
they would require qualities of an omniscient%
\footnote{Remember that relativity is retrodictive.} %
god ! For these reasons, we shall say that such works belong to {\em ontic relativity}.%
\footnote{From the Greek `ontos', {\em being}, with the meaning of `what is' as opposed to `how one knows obtains it'.}%

Ontic relativity is necessary to develop a relativistic experimental approach (and even sufficient in static, stationary, periodic or almost periodic gravitational fields!), but it is also generically {\em insufficient}. 

Let us remember that the statement of any problem in general relativity implies:
\begin{itemize}
\item[$\bullet$] to describe it directly in the space-time, by means of intrinsic relativistic objects, i.e. by means of worldlines and worldtubes (the analogs to the histories or evolutions of the instantaneous objects of Newtonian theory),
\item[$\bullet$] to banish generically the use of any physical extended present (due to high precision clocks or regions of non-constant inertial or gravitational fields). 
\end{itemize}
The works of relativity that, in addition:
\begin{itemize}
\item[$\bullet$] integrate the physicist as an element of the problem considered, 
\item[$\bullet$] concern physical properties that the physicist can measure, and
\item[$\bullet$] take into account explicitly what information, when and where, the physicist is able to know,
\end{itemize}
will be considered as characterizing {\em epistemic relativity}.%
\footnote{From the Greek `episteme', {\em knowledge}, with the meaning of `how we obtain it'.} %

Paradoxically, until now a very few number of relativistic problems have been solved explicitly under these assumptions.
%
%
\subsection{The ingredients of epistemic relativity}
The main objective of epistemic relativity is to provide the physicist with the {\em knowledge} and {\em protocols} necessary to make relativistic gravimetry (chronometry in arbitrary directions) in any {\em unknown} space-time environment.
\begin{figure}[h]
\begin{center}
\includegraphics[height=4cm]{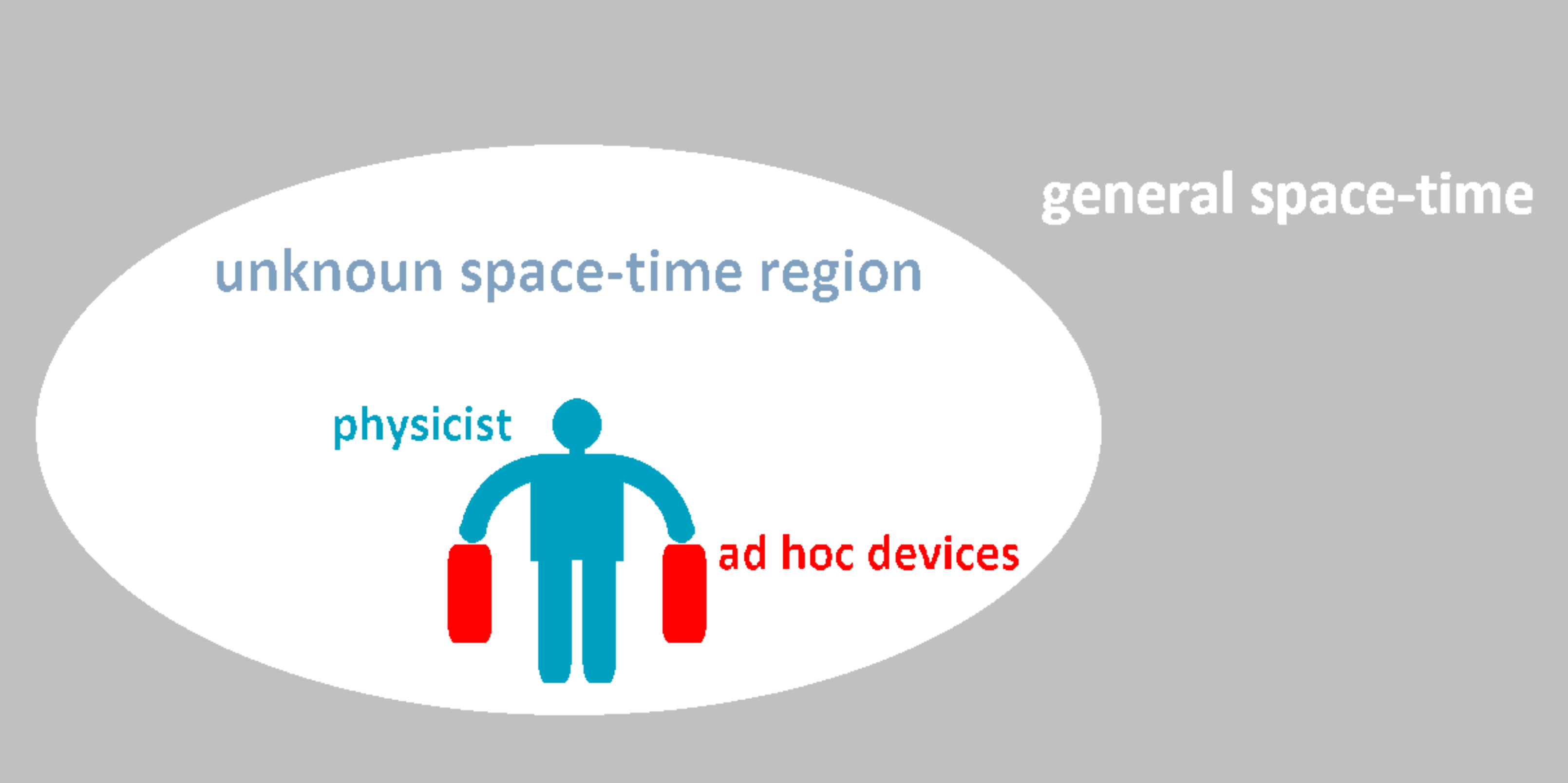}
\end{center}
\vspace{-3mm}
\caption{Epistemic relativity wants to provide the physicist with the baggage and protocols necessary to make gravimetry in any unknown region of arbitrary space-times.}
\label{baggage_fig:1}
\end{figure}

This is the first and unavoidable step to develop experimental relativity as the {\em natural scientific approach to the physical world}.

The basis of relativistic gravimetry are Einstein equations, and these equations need a precise mathematical model of the space-time. The adequacy of this mathematical model and the physical space-time that it describes implies a one-to-one correspondence between the points of the model and the physical events they describe. And because in the mathematical model points are precisely identified (by their coordinates), we need to know how to locate the events in the physical space-time.

What else do we need in epistemic relativity? In fact, what we need is to transform the space-time region of physical interest in a (finite) laboratory.

But, what is a laboratory in relativity? In fact, a reflection on this matter shows that any laboratory, regardless of the specificity of its measurement devices, has to provide us with: 
\begin{itemize}
\item[$\bullet$] a precise physical location of the significant parts of the system, and
\item[$\bullet$] a precise physical description of its intrinsic properties.
\end{itemize}

The devices able to carry out these two tasks are called {\em relativistic positioning 
systems} and {\em relativistic stereometric systems} respectively. Thus: 

\vspace{3mm}
A finite {\em laboratory} in relativity is a {\em space-time region} endowed with

\vspace{2.5mm}
\hspace{2cm} {$\bullet$ a relativistic positioning system and }

\hspace{2cm} {$\bullet$ a relativistic stereometric system.}
\vspace{3mm}

The first task of epistemic relativity is to transform the finite regions of interest of the space-time in laboratories or, equivalently, to construct these two relativistic systems.

These two relativistic systems are sufficient to describe any experiment in their region of space-time. Remember that, as already mentioned, we refer to physical systems as {\em relativistic objects}, i.e. as worldlines or worldtubes so that they naturally contain the changes of state that the Newtonian physical systems at every instant may suffer during an experiment (they already are the histories or evolutions of Newtonian systems). With the same point of view, the physical properties of relativistic objects cannot but be the {\em history} of the physical properties of the Newtonian objects at every instant of its constituent events. It is thus clear that the two above functions of locating and detecting are effectively able to describe any physical experiment in the space-time domain in question.
%
%
\section{Relativistic positioning systems}
%
%
\subsection{What is {\em to locate}}
In general, to {\em locate} an object is to {\em identify the place} it takes up. 
To locate with accuracy objects of any size (avoiding vagueness such as “near of”, “inside of”, etc.), we need to assign a proper name to every event of the space in question.
A {\em locating system} is a system of assignation of proper names to the events of a space.

In particular, because the space-time is a continuum, the assignation of a proper name to every one of its events has to be done with numbers. And because it is a four-dimensional continuum, every proper name has to be done with four numbers. 
So that, specifically, in (a region of) the space-time, a locating system is a system of assignation of four numbers to every one of its events.

In physics, the devices for assigning four numbers to every event may be constructed with different materials and be based in different physical properties, so that their specific construction and protocols of use give rise to different locating systems. 

The set of tetrads of numbers generated in a region by a locating system constitutes (a physical realization of) a coordinate system. On the other hand, in topological manifolds, a coordinate system in a region $\cal{R}$ may be extracted from a local chart $(\cal{R}, \varphi)$ by means of the coordinate functions associated to $\varphi$, where $\varphi$ is a homeomorphism from $\cal{R}$ to the associated linear space. We see that, mathematically, locating systems are represented by homeomorphisms $\varphi$. Thus, a locating system in a finite region $\cal{R}$ of the space-time may be seen as the physical construction of a local chart or, conversely, a local chart may be seen as a mathematical idealization of a locating system.%
\footnote{It is to be noted, because frequently forgotten, that the local charts on a differentiable manifold may be {\em structural}, i.e. belonging to the atlas defining its differentiable structure, or {\em not}. Only in the first case, the region $\cal{R}$ has to be an open set. Nonstructural local charts may be of lesser differentiable class than structural ones. In particular, they may be simply continuous, although in this case, the natural frame being absent, one may be lead to complete the local chart with an independent field of vector tetrads in order to construct a basis for the tensor algebra.} 
%
%
%
\subsection{Properties of locating systems}\label{subsecproplocsys}
According to the characteristics of the assignation system of coordinates, a locating system may be {\em active} if its assignation system only operates for events that emit signals of presence, or {\em passive} if it operates for all events of the region irrespective of their emission state; and it may be {\em immediate} if the values of the coordinates of every event are obtained without delay, or {\em retarded} otherwise. Also, according to its use, a locating system is said {\em real} if its assignation system starts one for all in all the region, and it is said {\em virtual} if it is used case by case to obtain only the specific coordinates of particular events. 

Another classification takes into account the {\em function} allocated to the locating system. Two locating systems are of particular importance for us: {\em reference systems}, which allow {\em one} observer to know the coordinates of the events of the region and {\em positioning systems}, which allow every event of the region to know its proper coordinates. 
\begin{figure}[h]
\begin{center}
\includegraphics[height=4cm]{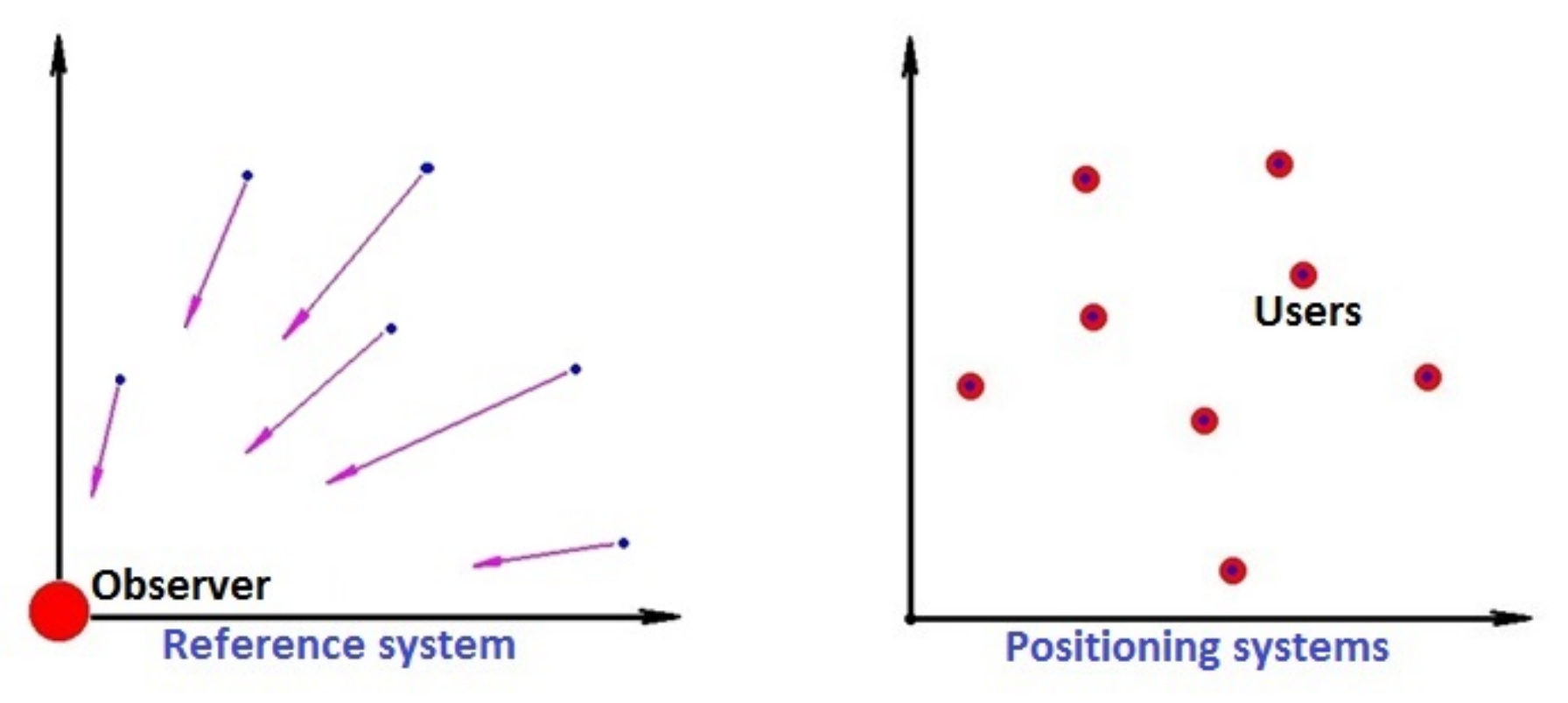}
\end{center}
\vspace{-4mm}
\caption{A reference system allows one observer to know the coordinates of the events of the region. A positioning system allows every event of the region to know its proper coordinates.}
\label{RefYPos_fig:2}
\end{figure}

It is then evident that real locating systems may be immediate and passive, meanwhile virtual locating systems are necessarily active and retarded. And, taking into account that in relativity information propagates at a finite velocity, it results that {\em all} reference systems are retarded.

We can conclude that, when it is possible to construct them, real positioning systems are the locating systems that offer the best physical performances.
%
%
\subsection{Relativistic positioning systems}
We have just seen that, among all locating systems, the real positioning systems offer the performances of being passive and immediate. These performances are still insufficient for us because the main objective of epistemic relativity is to make relativistic gravimetry in an {\em unknown} space-time environment.

Because the space-time environment is unknown, the locating system has also to be {\em generic}, that is to say, able to be constructed in regions of {\em any} space-time.%
\footnote{It is understood that such regions have the appropriate accessibility characteristics to make gravimetry.} %
And because the main objective is to make relativistic gravimetry, it has also to be (gravitationally) {\em free}, i.e. able to be constructed without the previous knowledge of the space-time metric. 

{\em Relativistic positioning systems} are the real, generic and free positioning systems of the space-time.

The construction of relativistic positioning systems is unexpectedly simple. A clock may be seen as a {\em continuous generator of numbers}, namely the time that it displays at every instant. Broadcasting this time by means of an electromagnetic signal, every event receiving the signal will know this number at the instant of reception. And because the proper name of every event of the space-time consists of four numbers (its coordinates), four clocks arbitrarily distributed will provide every event reached by the four signals with such numbers.
It is thus easy to see that: Four clocks broadcasting their times constitute a relativistic positioning system%
\footnote{Generically. Some distributions of clocks may associate same coordinates to different events, whatever the region considered (see \cite{Coll-Ferrando-Morales:2010a}).} %
(see \cite{Coll:2001}). 

The coordinates generated by relativistic positioning systems are called {\em emission coordinates} \cite{Coll:2006}. 

Note: I have chosen to introduce the well-known notion of coordinate system in this unusual form because the traditional presentations induce to give too importance to ingredients that are in fact dispensable or of secondary interest, but wrongly make many people to feel them necessary to understand physically a coordinate system. Thus, ingredients like {\em origin}, {\em coordinate lines} or {\em synchronization} of a coordinate system are respectively irrelevant, of secondary interest or inexistent%
\footnote{Or fourfold. Generically there does not exist a privileged synchronization of the form {\em} time = {\em constant} in relativistic positioning systems, but emission coordinates being constituted by four times $\tau^A$, one could say that there exist four synchronizations $\tau^A$ $= constant.$} %
for relativistic positioning systems. 

It is worthwhile to remark that relativistic positioning systems are a particular class of positioning systems but that, although similar in some aspects to positioning systems like GNSSs (global navigation satellite systems), differ clearly from them, whatever be the relativistic formalism with which these last ones are analyzed. These GNSSs are {\em technical} objects, wonder technical objects, with a lot of technical, social and even scientific applications. But they are not {\em scientific objects}. 
The reason is that they have been thought, calculated and constructed using consciously a defective theory, the Newtonian one, to construct a {\em Newtonian space} in the region between the Earth surface and the constellation of satellites, in which the International Atomic Time on the Earth surface is imposed on all the region as a Newtonian absolute time. The ``relativistic effects"%
\footnote{Euphemism for ``Newtonian defaults".}” %
are not used to improve the approximate Newtonian calculations and to obtain more precise values of the {\em physical} times involved, but to deform the values of these physical times so as to mimic an unphysical Newtonian time everywhere (the calculated ``relativistic effects" are {\em subtracted} to the physical values, not {\em added} to the Newtonian calculations of them!). 

The above comment is {\em not at all} a criticism of GNSSs: as technical objects, they have to be useful to us and give us what we ask them to give.
For relational (technical, social) convenience, it is the case, in particular, of a universal time for all of us, be we on the Earth surface or at 15.000 km height, at rest or at whatever velocity. But, I think that it would be better to have, first of all, a scientific object, an R-GNSS (Relativistic GNSS) able to give us our proper physical time, the value of the gravitational field or proper local distances, whatever our position and velocity with respect to the Earth surface. The reason is that starting from such a scientific object, a simple software would be able also to mimic GNSSs with their absolute time everywhere. The (theoretical) question is not so much to improve GNSSs, but instead to construct cutting-edge scientific R-GNSSs. 

There are no specific coordinates associated to GNSSs.%
\footnote{The pseudo-ranges are only considered as parameters able to calculate conventional coordinates on the 
Earth surface (WGS84, ITRS or others). The appellation `GPS coordinates' refers to these conventional coordinates provided by the GPS.} %
Nevertheless, emission coordinates are an important tool in the study of relativistic positioning systems and in particular in R-GNSS. Among others, they suggest new questions, provide new results and allow the constellation of satellites of an R-GNSS to generate {\em autonomous} positioning systems for the Earth.
%
%
\subsection{Auto-locating systems}
\begin{figure}[h]
\begin{center}
\includegraphics[height=5cm]{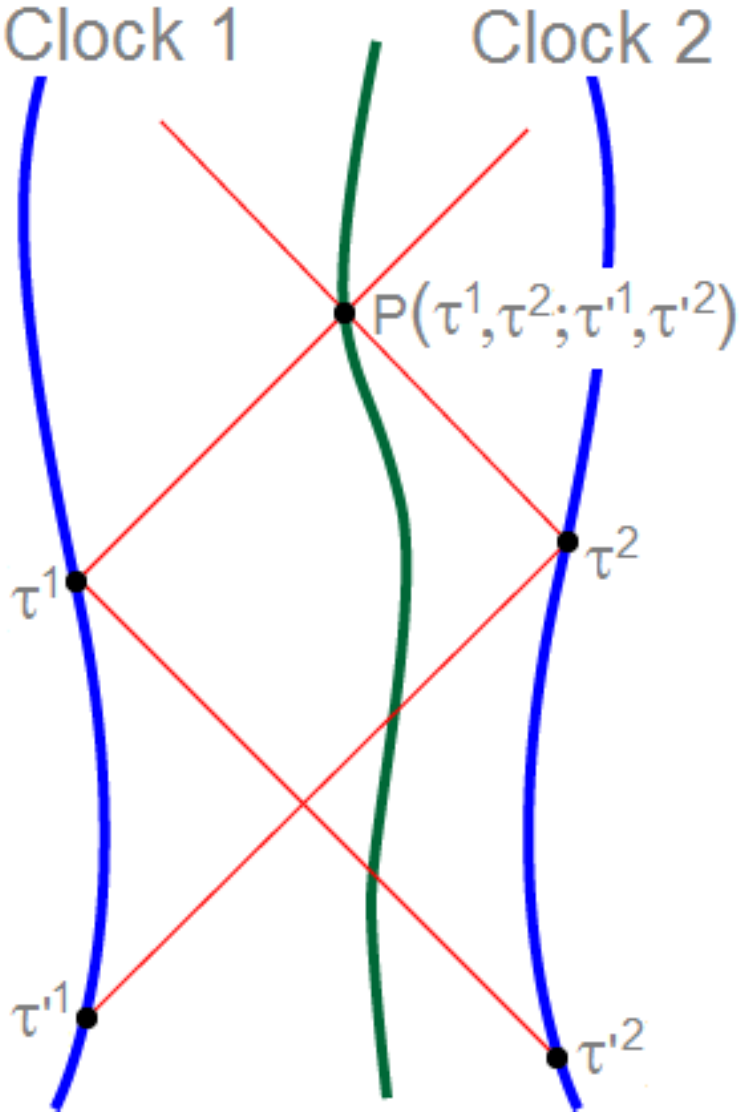}
\end{center}
\vspace{-4mm}
\caption{An auto-locating system allows any user to know its proper trajectory in emission coordinates, but also the trajectories of the clocks of the system (here a two-dimensional scheme).}
\label{2D_ALS_figc1}
\end{figure}

As an intermediate step, {\em auto-locating systems} are those relativistic positioning systems whose clocks are endowed with a transponder broadcasting the times they receive from the other clocks of the system (see \cite{Coll:2001},\cite{Coll-Ferrando-Morales:2010b} and references therein). So, every event $P$ of the region receives the sixteen times $\tau^A$ and $\tau^{AB}$ $A,B = 1,...,4, A \ne B$. The four $(\tau^A)$ are the emission coordinates of $P$, meanwhile every one of the four sets $A$ of four times $(\tau^A, \tau^{AB})$ are the emission coordinates of the clocks $A$ of the system (see Fig. \ref{2D_ALS_figc1}).
Thus, any user that records these data is able to know its trajectory as well as the trajectories of the clocks in the {\em grid} of emission coordinates.%
\footnote{ The grid $\cal G$ of emission coordinates is the Cartesian product of the segments $[\tau^A]$ of the times broadcast by the relativistic positioning system: $\cal G$ $\equiv$ $[\tau^1] \times [\tau^2] \times [\tau^3] \times [\tau^4]$.} %

An {\em echo-interval} for the clock $A$ due to the clock $B$ is the segment of world line of $A$ from the instant of emission of a broadcast signal by $A$ to the instant that it receives that signal echoed by $B$. In two dimensions, it is already proved a very interesting theorem for auto-locating systems (see \cite{Coll-Ferrando-Morales:2010b}): 

\vspace{3mm}
{\bf Theorem 3.4-1} {\em In a flat two-dimensional space-time, consider a user of an auto-locating system that knows its trajectory and those of the two clocks in the grid. If it also knows the acceleration of one of the clocks during one echo-interval, then it knows its acceleration and those of the two clocks along all their known trajectories}. 
\begin{figure}[h]
\begin{center}
\includegraphics[height=5cm]{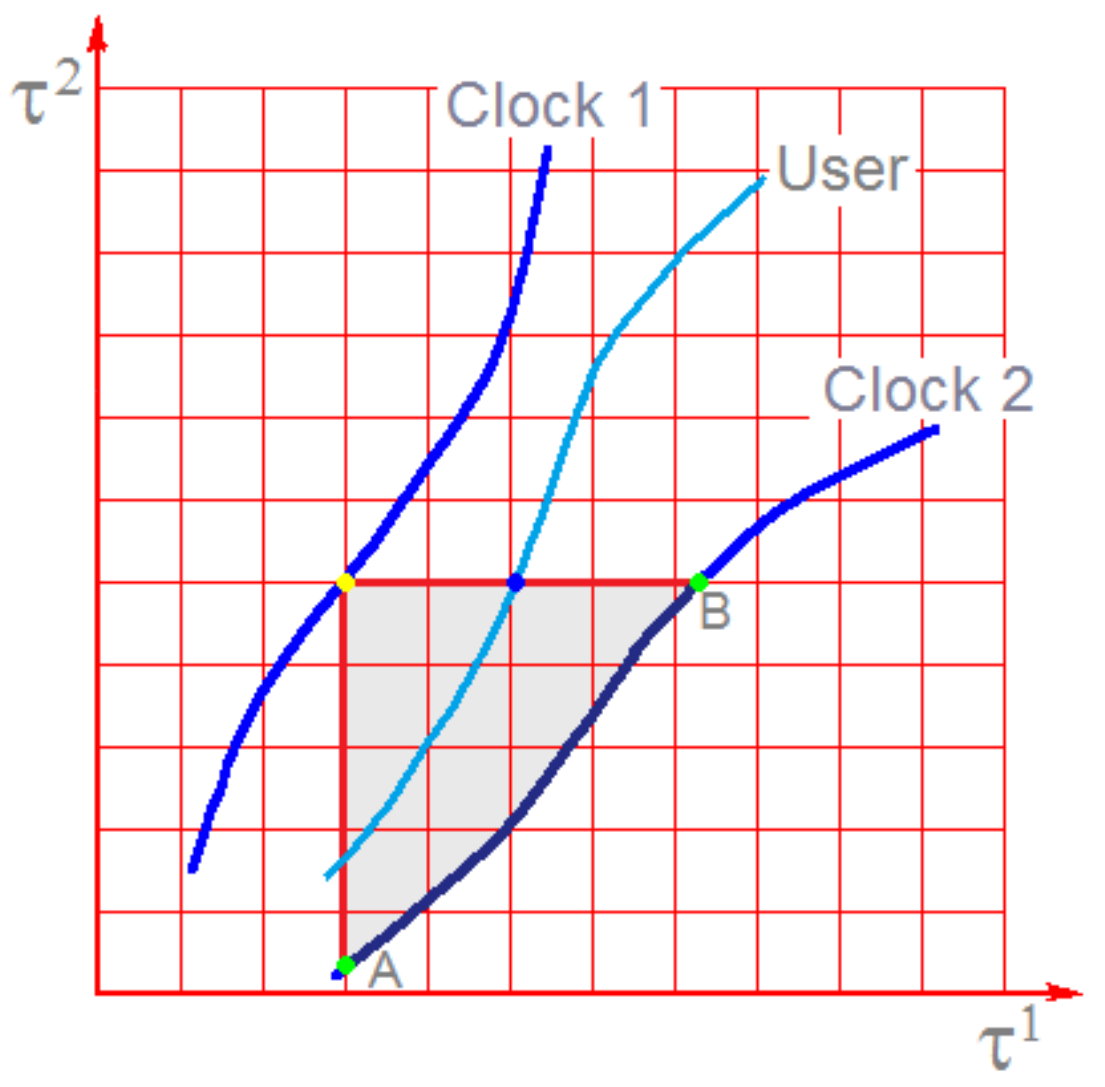}
\end{center}
\vspace{-4mm}
\caption{In absence of a gravitational field, a user of an auto-locating system is able to know its trajectory and that of the clocks in the grid of emission coordinates. If it also knows the acceleration of one of the clocks during one echo-interval AB, then it knows its acceleration and those of the two clocks along {\em all} their trajectories.}
\label{GridUserClocksEcho_fig}
\end{figure}

This theorem is very interesting because of the following fact: the algorithm that proves it in absence of a gravitational field may be as well applied in presence of it. Of course, in this case, it will provide a wrong acceleration along all the three trajectories excepting on the starting echo-interval of the clocks. But the importance of this fact is that the difference between these wrong accelerations (calculated as if the gravitational field were absent) and the true accelerations (measured by accelerometers along the world-lines of clocks and user in the gravitational field), is {\em all the dynamical effect that the gravitational field produces on the auto-locating system and the user}. 

Thus, it is a {\em complete}, although {\em relative} (to the choice of echo-interval), gravimetric measure.
Unfortunately, this fundamental theorem 3.4-1 is only know in two-dimensional space-times. It is of crucial importance for the theory of relativistic positioning systems and for all its near-future applications (Earth surface, RNGSS, Deep Space Navigation) to have its four-dimensional generalization. But this problem remains open. As in the two-dimensional case, the starting point is 
\begin{itemize}
\item[$\bullet$] the explicit knowledge of the coordinate transformation between the emission coordinates of four arbitrarily accelerated clocks and inertial coordinates, and
\item[$\bullet$] the explicit knowledge of the metric in emission coordinates.
\end{itemize}

Fortunately, the solution to these problems is already known. If the world-lines $\gamma_A(\tau^A)$, $A=1, ..., 4,$ of four emitters with respect to inertial coordinates are known, we can evaluate the quantities 
$$
e_a \equiv \gamma_a - \gamma_4 \quad , \quad \Omega_a \equiv \frac{1}{2}(e_a)^2 \quad , \quad \chi \equiv *(e_1 \wedge e_2 \wedge e_3 ) 
$$
$$ H \equiv *(\Omega_1\, e_2 \wedge e_3 + \Omega_2\, e_3 \wedge e_1 + \Omega_3\, e_1 \wedge e_2 ) \quad , \quad y_* \equiv \frac{1}{\xi . \chi}\, i(\xi)H
$$
where $\xi$ is an arbitrary transversal vector, $\xi .\chi \neq 0$, and $a = 1,2,3$. Then, the answer to the first item is (see \cite{Coll-Ferrando-Morales:2010a} and \cite{Coll-Ferrando-Morales:2012}):
\vspace{3mm}

{\bf Theorem 3.4-2} {\em Suppose known the world-lines $x = \gamma_A(\tau^A)$ of the four emitters of a positioning system with respect to an inertial coordinate system $\{x\}$, and let $\{\tau^A\}$ be their emission coordinates. In term of the quantities $\chi$ and $y_*$ evaluated from the $\gamma_A(\tau^A)$'s, the coordinate transformation $x = \kappa(\tau^A)$ between emission and inertial coordinates is given by:
$$
x \equiv \kappa(\tau^R) = \gamma_4 + y_* + \frac{y_*^2\,\chi}{(y_*\, .\,\chi) +\hat\varepsilon\sqrt{(y_*\,.\,\chi)^2 -y_*^2 \,\chi^2}}
$$
where $\hat\varepsilon$ is the orientation of the position system with respect to the user.}
\vspace{3mm}

And the answer to the second item is:

\vspace{3mm}
{\bf Theorem 3.4-3} {\em In terms of the above transformation $x = \kappa(\tau^R)$ and the world-lines $x = \gamma_A(\tau^A)$, the contravariant components of the metric, $g^{AB}(\tau^R)$, in the emission coordinates $\{\tau^R\}$ are given by:
$$
g^{AB}(\tau^R) = \frac{\Omega_{AB}}{\mu_A\,\mu_B} \quad ,
$$
where
$$
\Omega_{AB} \equiv \frac{1}{2}(\gamma_A - \gamma_B)^2 \quad , \quad \mu_A \equiv (\kappa - \gamma_A)\, . \, \dot{\gamma}_A \quad .
$$
}
%
%
\subsection{Autonomous positioning systems}
Auto-locating systems contain {\em all} what a coordinate system needs in order to be drawn in a space in which a field of light-cones is given.%
\footnote{The field of light-cones defines all the possible coordinate hypersurfaces of the coordinate systems associated to all the possible relativistic positioning systems. Accordingly, the particular trajectories of the clocks select the specific ones for the auto-locating system and, consequently, its region of validity or domain of definition in the grid.} %
But if the light-cones are specifically related to a geometry (here the gravitational field represented by a metric), and we want to identify individually the coordinate system making in it gravimetry (measure of the metric), we need additional information. This information has to be also broadcast by the clocks and consists of:
\begin{itemize} 
\item[$\bullet$] dynamical data of the satellites (acceleration, gradiometry),
\item[$\bullet$] observational data from them (e.g. particular masses, directions of reference quasars or pulsars, etc.), and
\item[$\bullet$] gravitational knowledge of the coordinate region (theoretical, experimental or mixed).
\end{itemize}
This information constitutes the {\em autonomous data}. {\em Autonomous positioning systems} are auto-locating systems that broadcast autonomous data. 

Autonomous positioning systems are, generically, the best locating systems we are able to construct. Their notion was already proposed in the first paper on positioning systems (see \cite{Coll:2001})
\section{Relativistic stereometric systems}
%
%
\subsection{The notion}
We already know that, to transform a finite region of the space-time in a laboratory, we need to endow it with, on one hand, a relativistic positioning system, to locate in the best form the significant parts of the physical system in question, and on the other hand, a relativistic stereometric system, to obtain their intrinsic physical properties. 

Similarly to a relativistic positioning system, that consists essentially of four clocks broadcasting their times,%
\footnote{The time broadcasted by every clock may be any convenient time, not necessarily its proper time, although proper time simplifies, in general, the theoretical analysis.} %
{\em a relativistic stereometric system consists of a system of four observers receiving the signals broadcast from a physical system.}

In relativity, the word `observer' is used in different contexts with different meanings. Here an {\em observer} means an {\em eye} able to record and to analyze the input, and an eye%
\footnote{More precisely, a $4\pi$-wide hypergon eye or $4\pi$ steradian eye.} %
is a small (physically {\em local}, mathematically {\em differential}) object, determined by its space-time worldline, able to project its past light-cone on its celestial sphere.

\begin{figure}[h]
\begin{center}
\includegraphics[height=3cm]{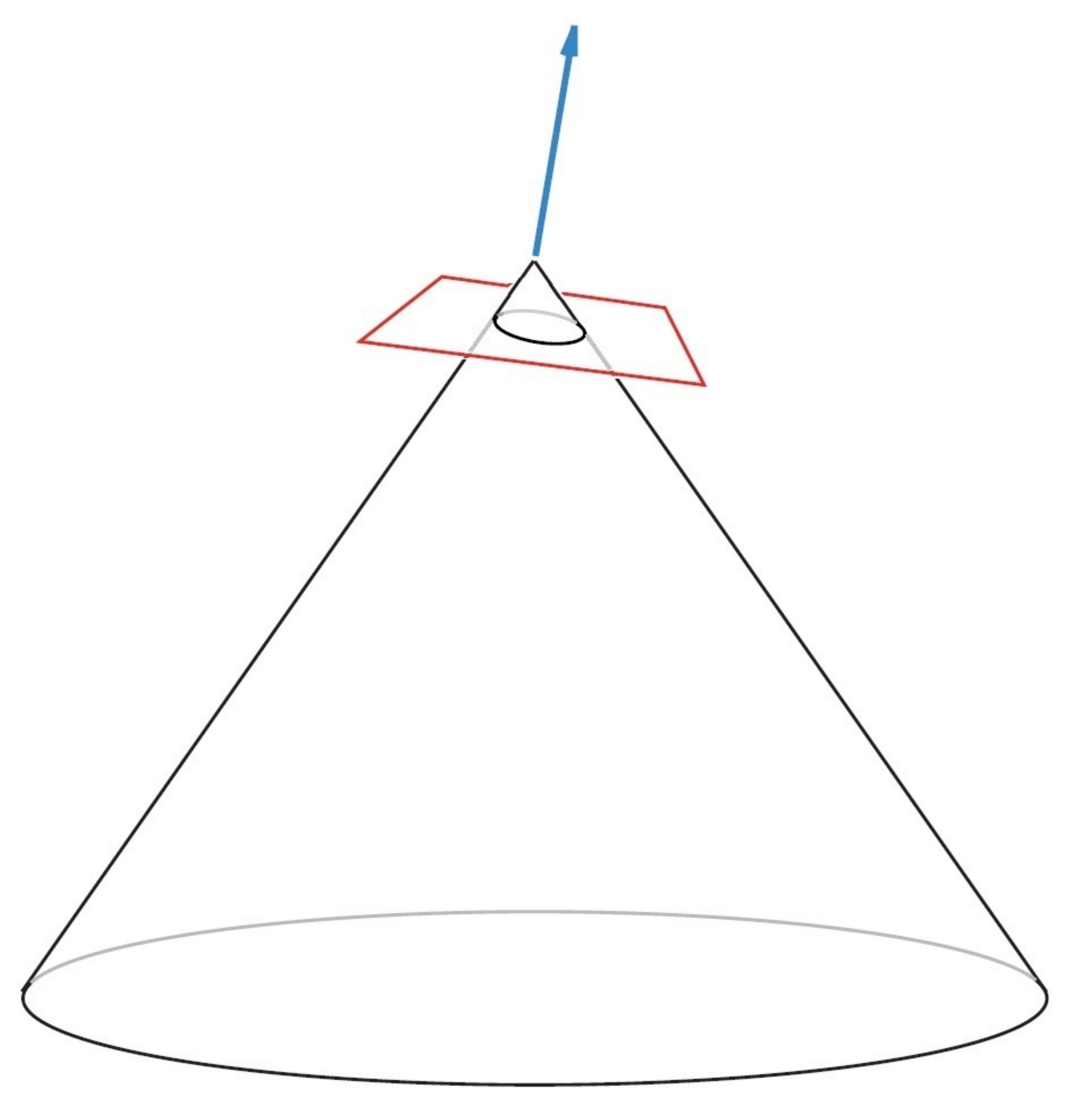}
\end{center}
\vspace{-4mm}
\caption{An observer, determined by its unit velocity, projects its past light-cone on its celestial sphere.}
\label{Ojo_fig:3}
\end{figure}
There exist interesting papers on relativistic vision but they are very different in strategy, starting hypothesis and definition of `eye', and their number is very small. It would be worthwhile to analyze and classify them, and especially to select those related to hypergon eyes.
 
But almost none of them consider the invariants of the observed figures. The study of these invariants is important for positioning as well as for stereometry. For example, any set of four events in the space-time (four stars in the sky or four material points of a system), that are seen on a circle of the celestial sphere of an observer, will be seen also on a circle, by any other observer at the same event, whatever be their relative velocity at the event. In addition, the relative velocity of such two observers may be obtained by comparison of the relative positions on the circle of the four points as seen by them.%
\footnote{In principle, a suitable choice of stars, in e.g. Hipparcos star catalog, could help a spacecraft in the Solar system to estimate its velocity with respect to the Barycentric Celestial Reference System.} %
\begin{figure}[h]
\begin{center}
\includegraphics[height=3cm]{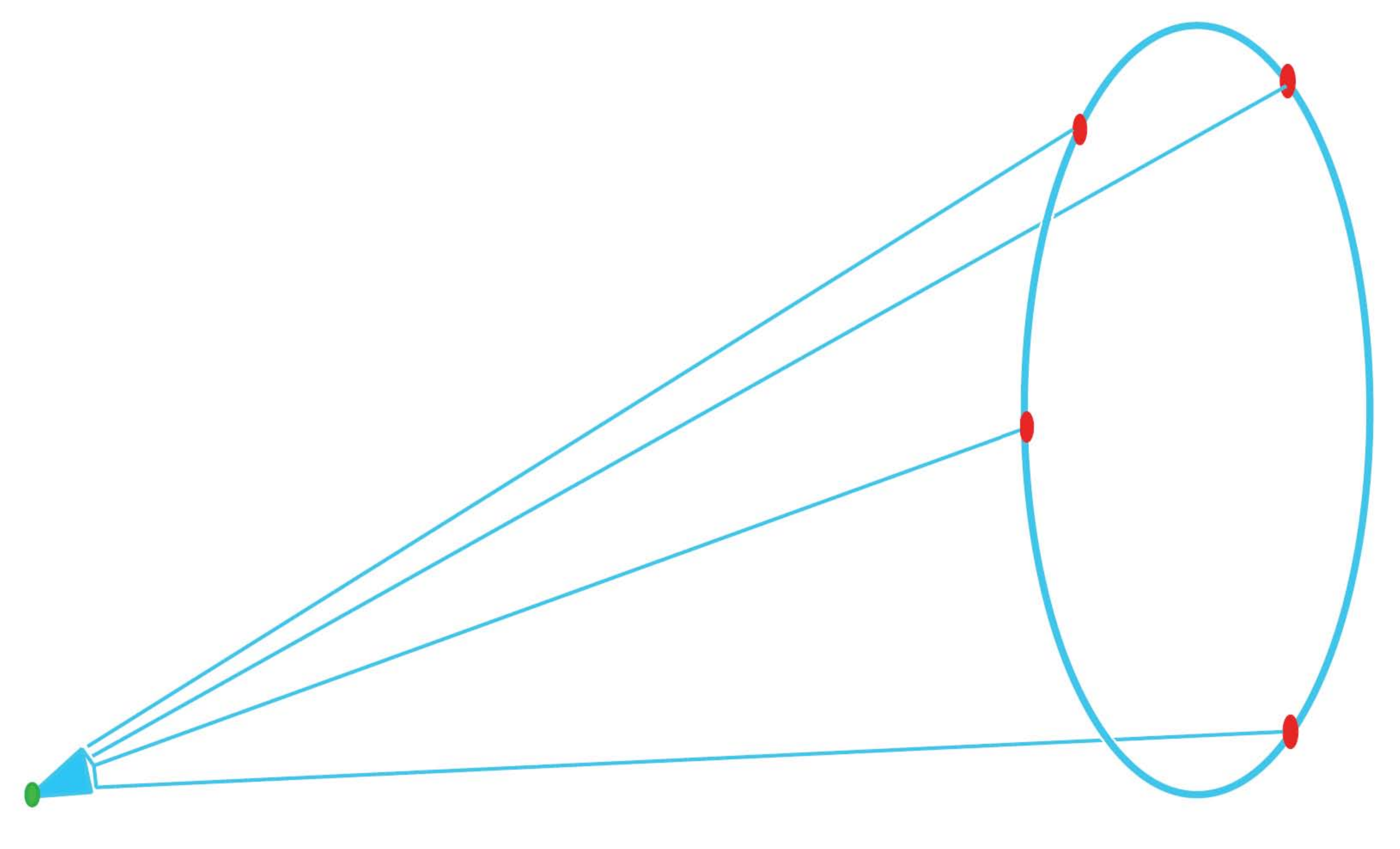}
\end{center}
\vspace{-4mm}
\caption{Four events that are seen in a circle on the celestial sphere of an observer are also seen in a circle on the celestial sphere of any other observer at the same event, whatever its relative velocity.}
\label{Ojo4Estrellas_fig:4}
\end{figure}

From the point of view of the space-time geometry, {\em relativistic stereometric systems are the causal duals of relativistic positioning systems}: the four observers ({\em resp.} four clocks) of the relativistic stereometric system ({\em resp.} relativistic positioning system) receive ({\em resp.} broadcast) the signals emitted by the event to be analyzed
({\em resp.} the signals received by the event to be located). As a consequence, every geometric result on one of these systems generates another geometric result for the other system, by a simple change of time orientation. 
\begin{figure}[h]
\begin{center}
\includegraphics[height=3cm]{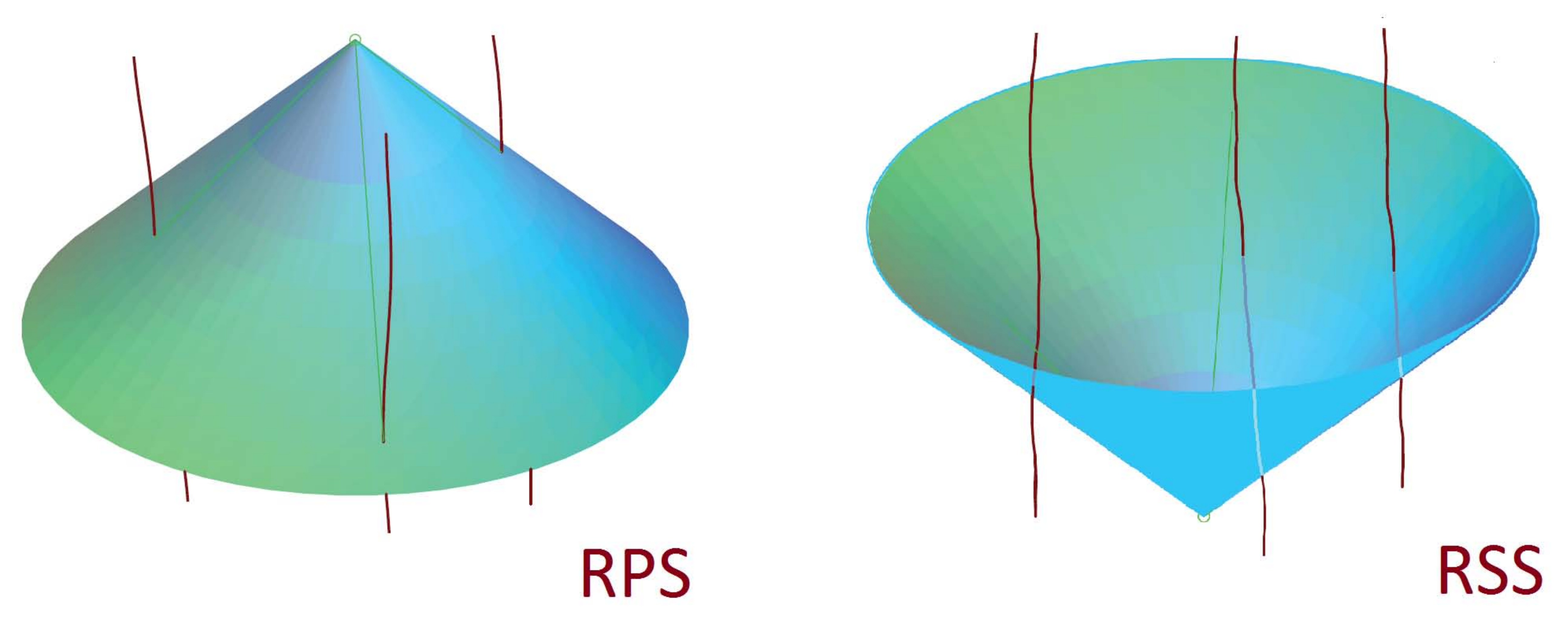}
\end{center}
\vspace{-4mm}
\caption{Relativistic stereometric systems are the causal duals of relativistic positioning systems.}
\label{RPS&RSS_fig:5}
\end{figure}

Relativistic stereometric systems are conceived to obtain the intrinsic properties of physical systems but, in their basic structure, they are also locating systems. As locating systems, the coordinates of every event are the (four) times of reception by the (four) observers of the signal emitted by the physical system at this event. According to the properties considered in subsection \ref{subsecproplocsys}, relativistic stereometric systems are {\em active}, {\em retarded} and {\em virtual} locating systems, in contrast with the {\em passive}, {\em immediate} and {\em real} properties of relativistic positioning systems. As we have seen in subsection \ref{subsecproplocsys}, the active, retarded and virtual properties of relativistic stereometric systems are not the best as locating systems but, nevertheless, they are well adapted to the principal function of relativistic stereometric systems, because the intrinsic properties of a physical system cannot be obtained but by active, retarded and virtual ways. Thus, relativistic stereometric systems not only describe the intrinsic properties of events but also locate them.%
\footnote{Of course, by broadcasting their proper times, a relativistic stereometric system may also act as a relativistic positioning system. But because of their dual ways of working, I believe it is clearer, for the moment, to study them separately.}%
%
%
%
\subsection{First results in relativistic stereometry}
Every observer of a relativistic stereometric system receives a (relativistic) perspective of the properties broadcast by the events. These four {\em observer-dependent} perspectives have to be combined in order to obtain the
{\em observer-invariant} ({\em proper} or {\em intrinsic}) properties that generate them. 

To show how to proceed, we shall present here the simplest case of relativistic stereometry, that of a material point broadcasting a signal in a flat two-dimensional space-time. In spite of its simplicity: (i) it leads to a simple but interesting extension of the unidirectional Doppler formula and (ii) it shows the difference between an ontic law and an epistemic one.
\begin{figure}[h]
\begin{center}
\includegraphics[height=5cm]{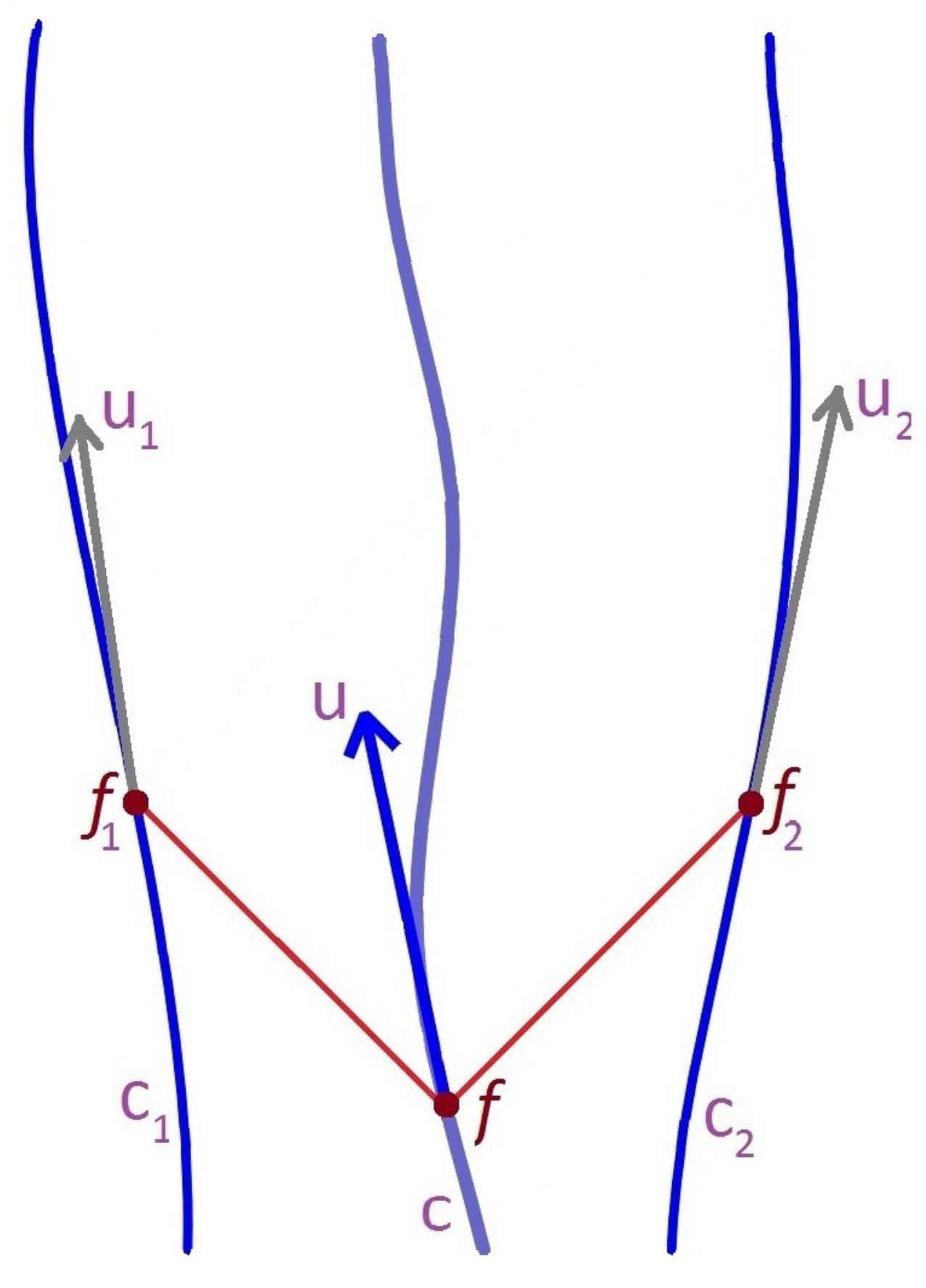}
\end{center}
\vspace{-4mm}
\caption{A two-dimensional relativistic stereometric system measuring the proper frequency $f$ of a
material point.}
\label{2DStereo_fig:6}
\end{figure}

Thus, let $C$, $C_1$ and $C_2$ be respectively the worldlines of a material point and of the two observers of the relativistic stereometric system in Minkowski space $M_2$. Suppose that at a particular instant%
\footnote{There is no matter here what instant-identifier is used: a clock associate to the point, measuring any time, non-necessarily proper, a flash, or any other pertinent one.} 
$C$ broadcasts a signal of proper frequency $f$ that is received by $C_1$ and $C_2$ at the Doppler frequencies $f_1$ and $f_2$ respectively. Let $v_1$ and $v_2$ be the scalar relative velocities of $C$ with respect to $C_1$ and $C_2$ evaluated by them at the instants of reception of the frequencies $f_1$ and $f_2$ respectively. As it is well known, $f_1$ and $f_2$ are related to $f$ by the Doppler expressions:
\begin{equation}\label{standarddoppler}
f_1 = f \sqrt{\frac{1 - v_1}{1 + v_1}} \quad , \quad 
f_2 = f \sqrt{\frac{1 - v_2}{1 + v_2}} \quad .
\end{equation}
Denoting by $v_{12}$ the relative scalar velocity of $C_1$ and $C_2$ at the instants of reception of the Doppler frequencies, one can prove:

\vspace{3mm}
{\bf Theorem 4.2-1.-} {\em In terms of the relative scalar velocity $v_{12}$ of the observers $C_1$ and $C_2$ of a relativistic stereometric system at the instant of reception of the Doppler frequencies $f_1$ and $f_2$ from a material point $C$, the proper frequency $f$ of the material point at the instant of emission is given by:} 
\begin{equation}\label{f1f2f}
f^2 = f_1f_2 \sqrt{\frac{1 + v_{12}}{1 - v_{12}}} \quad .
\end{equation}

\vspace{3mm}
Of course, if the material point is dragged by one of the observers of the system, say $f = f_2$, expression (\ref{f1f2f}) coincides with the first of (\ref{standarddoppler}).

The Doppler frequencies also allow to know the relative velocities of the material point $C$ at the instant of emission of the signal with respect to the observers $C_1$ and $C_2$ at the instants of its reception. By elimination of $f$ from every one of the expressions (\ref{standarddoppler}) and (\ref{f1f2f}), one has:

\vspace{3mm}
{\bf Theorem 4.2-2.-} {\em The relative velocities $v_1$ and $v_2$ of the material point $C$ at the instant of emission of the signal with respect to the observers $C_1$ and $C_2$ at the instants of its reception are given by:}
\begin{equation}\label{RelativeVelocities} 
v_1 = \frac{f_2\sqrt{1 + v_{12}} - f_1 \sqrt{1 - v_{12}}}{f_2\sqrt{1 + v_{12}} + f_1 \sqrt{1 - v_{12}}} \quad ,
\end{equation}
\begin{equation}
v_2 = \frac{f_1\sqrt{1 + v_{12}} - f_2 \sqrt{1 - v_{12}}}{f_1\sqrt{1 + v_{12}} + f_2 \sqrt{1 - v_{12}}} \quad .
\end{equation}

\begin{figure}[h]\label{D2StereoTaus_fig:7}
\begin{center}
\includegraphics[height=5cm]{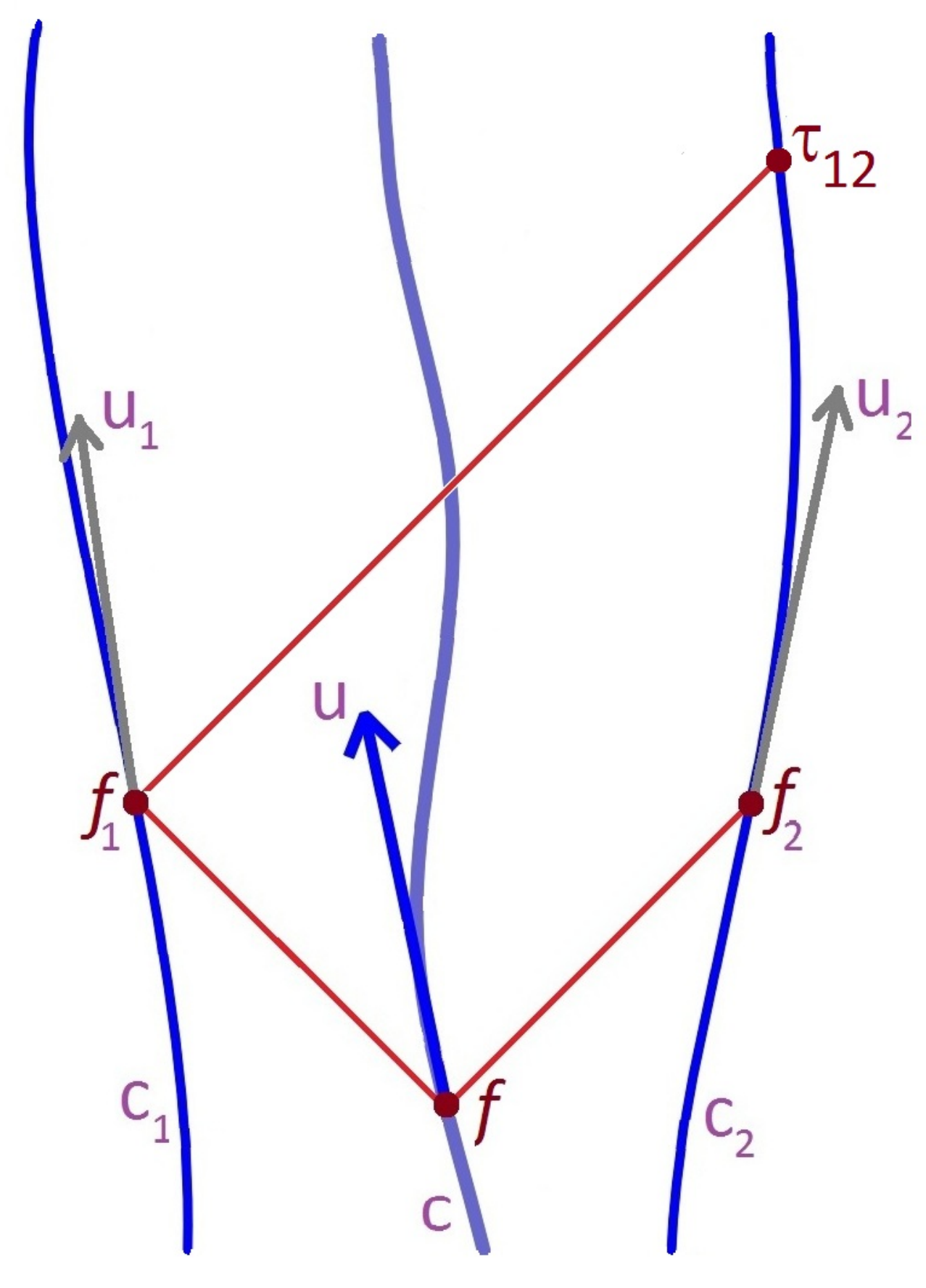}
\end{center}
\vspace{-4mm}
\caption{The observer $C_2$ has been chosen as the `physicist'. From the instant $\tau_{12}$ on, it may know all the quantities of the problem.}
\label{2DStereoTaus_fig:7}
\end{figure}

\vspace{3mm}
The above results depend on the Doppler frequencies $f_1$ and $f_2$ but also on the quantity $v_{12}$, the scalar velocity of one of the observers of the relativistic stereometric system with respect to the other at the instants at which they receive the Doppler frequencies. This quantity cannot be measured directly, but has to be calculated in terms of other physical quantities. This reason is sufficient to show that the above two theorems, in their present form, do not constitute results of epistemic relativity. In fact, these two theorems do not fulfill any of the three conditions of subsection 2.1 characterizing
epistemic relativity. For that, we must specify (i) what physicist we have chosen to make the experiment, (ii) when and where it will know the quantities needed to solve the problem and (iii) how can it know or measure these quantities.
In the present case, we make the following choices:
\begin{itemize}
\item[$\bullet$] the physicist is one of the observers of the relativistic stereometric system, say $C_2$, as shown in Fig. \ref{2DStereoTaus_fig:7},
\item[$\bullet$] it will be informed of the pertinent quantities at the instant $\tau_{12}$ of reception of the information coming from the observer $C_1$,
\item[$\bullet$] at that instant $\tau_{12}$, it already knows the quantity $f_2$ measured and recorded by itself, it is being informed of the quantity $f_1$, and then may know the quantity $v_{12}$ by computation. 
\end{itemize}

At the instant of reception of the signal $f_1$, the observer $C_1$, in addition to send its measure, must send also its reflection or its proper frequency so as to allow $C_2$ to know its relative velocity at its instant $\tau_{12}$. On the other hand, $C_2$ is supposed to know its proper acceleration, so that it is able to know its relative velocity between its instant of reception of the signal $f_2$ and the instant $\tau_{12}$. The knowledge of these two velocities, allow it to calculate the velocity $v_{12}$ of the theorems. For simplicity, suppose here that $C_2$ is geodesic. Then, the relative frequency between the proper time of $C_1$ at the instant of reception of the frequency $f_1$ and the proper time of $C_2$ at any instant is constant. Denoting it by $\nu_{12}$ and inverting the Doppler formula we have:
\begin{equation}\label{RelativeVelocity} 
v_{12} = \frac{1 - \nu_{12}^2}{1 + \nu_{12}^2} \; . 
\end{equation}

Now if, for short, we call an epistemic theorem an `epistem', the above two theorems in this geodesic case become respectively:

\vspace{3mm}
{\bf Epistem 1.-} {\em In terms of the frequencies $f_1$ and $f_2$ received by a relativistic stereometric system $\{C_1 , C_2\}$ from a material point $C$ and of the relative frequency $\nu_{12}$ of the proper time of the observer $C_1$ with respect to the geodesic observer $C_2$, the proper frequency $f$ of the material point $C$ is given by:}
\begin{equation}\label{episten1} 
f^2 = \frac{f_1 f_2}{ \nu_{12}} \; .
\end{equation}

\vspace{3mm}
{\bf Epistem 2.-} {\em In terms of the frequencies $f_1$ and $f_2$ received by a relativistic stereometric system $\{C_1 , C_2\}$ from a material point emitter $C$ and of the relative frequency $\nu_{12}$ of the proper time of the observer $C_1$ with respect to the geodesic observer $C_2$, the relative velocities $v_1$ and $v_2$ of the material point $C$ with respect to the observers $C_1$ and $C_2$ at the instants of reception of the signals $f_1$ and $f_2$ are given by:} 
\begin{equation}\label{epistem2} 
v_1 = \frac{f_2 - f_1 \nu_{12}}{f_2 + f_1 \nu_{12}} \quad , \quad v_2 = \frac{f_1 - f_2 \nu_{12}}{f_1 + f_2\nu_{12}} \, .
\end{equation}

\vspace{3mm}
These results are very simple, but show roughly the way of working with stereometric systems in epistemic relativity. Of course, the problem of determining the proper distances to its neighboring elements of a material point cannot be considered in a two-dimensional space-time, because there the celestial sphere of an observer reduces to two opposite points. It is evident that this work has to be extended and generalized in three and four dimensions. It remains an open problem.
%
%
\section{Intrinsic characterization of gravitational fields} \label{Intrinsic}
In epistemic relativity, events are located in emission coordinates generated by relativistic positioning systems.

A complete set of gravimetric measurements (whatever be the methods) will lead to the experimental values of the components of the metric in emission coordinates. 

It rests to identify this metric or, equivalently, its sources. For example, suppose we suspect that it corresponds either to a Kerr or to a Schwarzschild gravitational field, but that it can be neither one nor the other. How can we discern between these three possibilities?

The most part of the known gravitational fields (exact or approximate solutions to Einstein equations), and in particular Kerr and Schwarzschild, are known in very particular coordinate systems which have no simple relation with emission coordinates. 

For example, for the Schwarzschild case, the direct procedure would be%
\begin{itemize}
\item[$\bullet$] to calculate in Schwarzschild coordinates the equations of the field of light-cones,
\item[$\bullet$] to model in these coordinates the world-lines of the clocks of the positioning system, parameterized with their proper time, 
\item[$\bullet$] to select with both results the four families of light-cones emitted by the clocks or, in other words, to obtain the transformation from emission coordinates to the Schwarzschild ones, 
\item[$\bullet$] to solve for the inverse transformation, 
\item[$\bullet$] to transform the Schwarzschild metric components in emission metric components, 
\item[$\bullet$] to compare them with the experimental values obtained in these emission coordinates. 
\end{itemize}

Every one of these items presents non-negligible difficulties and can introduce specific uncertainties. It results in a hard and long procedure.

For this reason, to compare the experimental value of the gravitational field obtained in emission coordinates with a given solution of Einstein equations, it may be better to characterize intrinsically %
%
%
the solution and to check if the experimental data verify this characterization. 

By {\em intrinsic characterization of a metric} we understand here a set of necessary and sufficient local conditions on some of the concomitants of the metric that characterize it, its sources and their position, regardless of the coordinate system.

The idea is that an observer, from the sole measure of the gravitational field and its variations in a local region, is able to know the masses that produce it as well as its positions with respect to it.

\begin{figure}[h]\label{massYspacecraft_fig}
\begin{center}
\includegraphics[height=4cm]{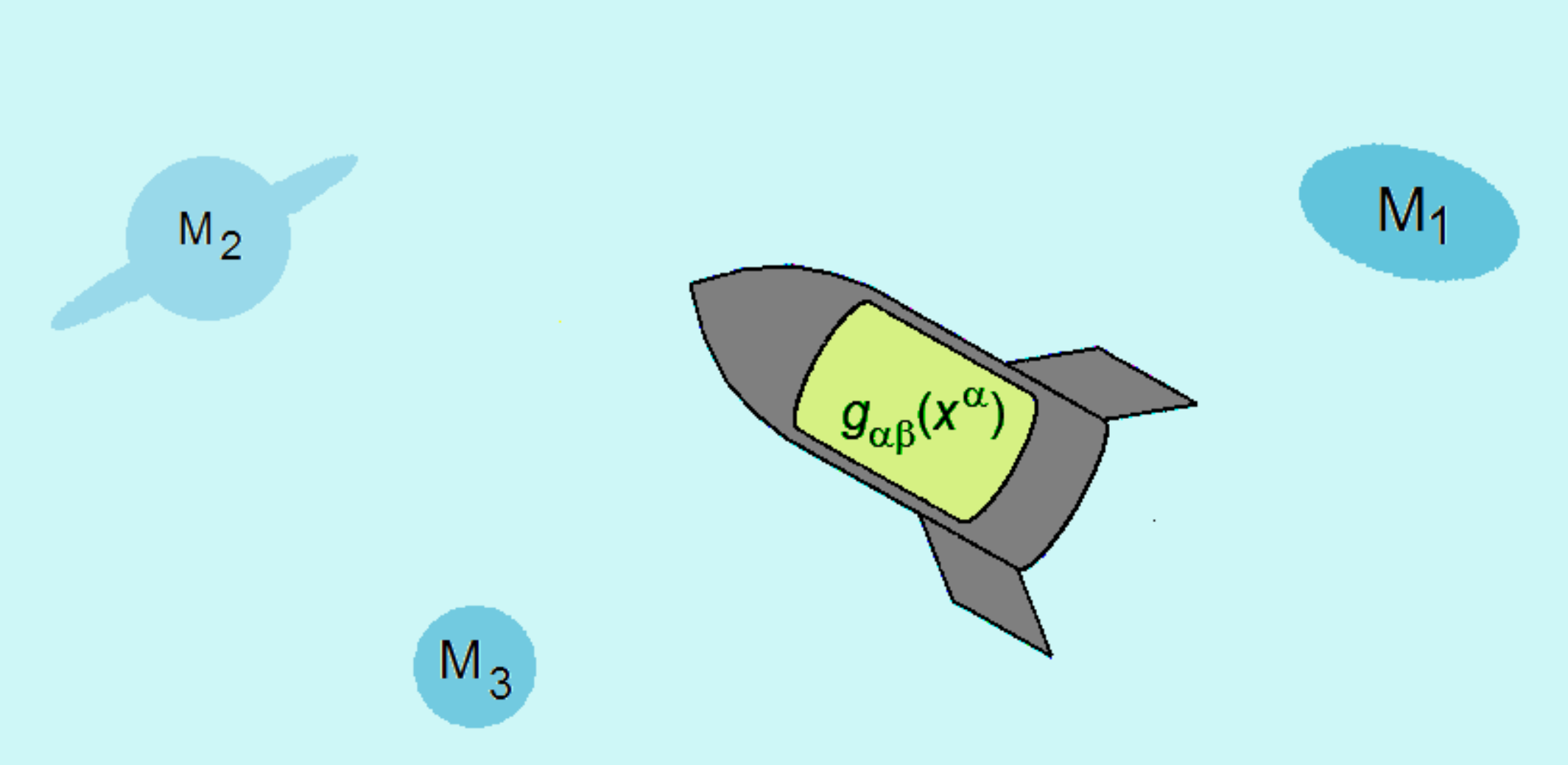}
\end{center}
\vspace{-4mm}
\caption{An observer that measures the gravitational field in a local space is able to know the masses that produce them as well as their positions.}
\label{massYspacecraft_fig:7}
\end{figure}
The intrinsic characterization of individual metrics is, apart from some trivial cases, a relatively recent problem. I would like to draw attention on the pertinence not only of its development for all exact solutions of Einstein equations of physical interest but also of its extension to the usual approximate solutions in experimental applications. By the reason above 
indicated, I believe this characterization to be the simplest and shortest way to identify gravitational fields in epistemic relativity.

The idea to characterize individual metrics intrinsically originates some years ago as part of a general project of IDEAL solution of problems, somehow similar in spirit to that of classic Greek mathematicians but of vivid actuality for the problems we were concerned, where the acronym stands for
\begin{itemize}
\item[$\diamond$] {\bf I}ntrinsic (depending only on the concepts mentioned in the statement of the problem), 
\item[$\diamond$] {\bf D}eductive (not involving inductive or inferential methods or arguments),
\item[$\diamond$] {\bf E}xplicit (expressing the elements of the solution not implicitly) and
\item[$\diamond$] {\bf Al}gorithmic (giving the solution as a flow chart with a finite number of steps).
\end{itemize}

The first solution intentionally obtained under the IDEAL spirit was that of the point particle in Newtonian gravity (\cite{Coll-Ferrando:1997}). I cite it here because the solution to its relativistic analog, the Schwarzschild gravitational field, in contrast with an extended opinion but like many other generic problems,%
\footnote{Formulation of Maxwell equations, Cauchy problem for the permanence of electromagnetic waves, shock, detonation and deflagration waves in hydrodynamics, and some others.} %
admits simpler expressions than that of Newtonian physics.

The intrinsic characterization of the Schwarzschild gravitational field is due to J. Ferrando and J.A. S\'aez (\cite{Ferrando-Saez:1998}).
Denoting by $Riem$ the Riemann tensor, by $tr$ and $i(.)$ the trace operator and the interior product respectively, and by $\wedge$ the exterior product (considering the metric $g$ as a double 1-form), they introduce the two scalars:
\begin{equation}\label{alfaYsigma} 
\sigma \equiv \left( \frac{1}{12} tr^2Riem^3\right)^\frac{1}{3} \quad , \quad \alpha \equiv \frac{1}{9\,\sigma^2}\, g(d\sigma,d\sigma) + 2\sigma \quad, 
\end{equation}
allowing to construct the two tensors:
\begin{equation}\label{SandQ} 
S \equiv - \frac{1}{3\,\sigma}\left( Riem + \frac{1}{2}\sigma g\wedge g\right) \quad , \quad Q \equiv i^2(d\sigma)S \quad .
\end{equation}
and, with the help of an arbitrary direction $x$, generate a third, direction-dependent scalar $\theta$:
\begin{equation}\label{thetadef} 
\theta(x) \equiv 2Q(x,x) + tr Q \quad .
\end{equation}
These quantities may be evaluated for any gravitational field, and involve the gradient of a function of $Riem$, a third variation of the metric.

Their first result is: 

\vspace{3mm}
{\bf Theorem 5-1.-} {\em A vacuum metric $g$, $Ric(g) = 0$, is the Schwarzschild metric if, and only if, the three scalars $\sigma$, $\alpha$ and $\theta$ are strictly positive, 
\begin{equation}\label{scalarsIneq} 
\sigma > 0 \quad , \quad \alpha > 0 \quad , \quad \theta(x) > 0 \quad , 
\end{equation}
and the two following tensors vanish:
\begin{equation}\label{*riemandSeqs} 
i^2(d\sigma) \ast\!\! Riem = 0 \quad , \quad S^2 + S = 0 \quad , 
\end{equation}
where $\ast$ is the Hodge duality operator.}
\vspace{3mm}

In the circumstances suggested by Fig. 10 
this result assures the experimental physicist, whatever the coordinates used, that it is immersed in the field of a spherically symmetric mass. But what mass and where? The answer is also given in \cite{Ferrando-Saez:1998}: 

\vspace{3mm}
{\bf Theorem 5-2.-} {\em The mass $m$ and the radial coordinate $r$ of a Schwarz-schild metric are given by:}
\begin{equation}\label{teo5.2} 
m = \sigma \alpha^{-\frac{3}{2}} \quad , \quad r = \alpha^{-\frac{1}{2}} \quad . 
\end{equation}

In fact, any other intrinsic characteristic of Schwarzschild metric may be obtained in an IDEAL way. For example, differentiating the second of relations (\ref{teo5.2}) one obtains the intrinsic expression in terms of the invariant $\alpha$ of the radial codirection $-dr$ of the mass, allowing the physicist to know its direction, distance and value. Also, on the event horizon $r = 2m$ one would find $\alpha$ $= 2\sigma$ and, as (\ref{alfaYsigma}) shows, also $g(d\sigma,d\sigma)$ would vanish. Furthermore, as proven also in \cite{Ferrando-Saez:1998}, the time-like direction of static evolution of the gravitational field, i.e. its integrable Killing vector field $\xi$, is given by:
\begin{equation}\label{} 
\xi = - \sigma^{-\frac{4}{3}}\frac{Q(x)}{\sqrt{Q(x,x)}} \quad . 
\end{equation}

These results already allow us to appreciate the interest of local intrinsic characterizations of gravitational fields. But they are far from being {\em only} the characterization of the simplest gravitational field. The scalars $\sigma$, $\alpha$ and $\theta$, in (\ref{alfaYsigma}) and (\ref{thetadef}), and the tensors $S$ and $Q$, in (\ref{SandQ}), may be evaluated in {\em any} gravitational field and their comparison with relations (\ref{scalarsIneq}) and (\ref{*riemandSeqs}) will give us an information that we would study in detail, in particular if the gravitational field in question may be well described as a Schwarzschild perturbation. 

At present, the intrinsic characterization of some other classes of vacuum exact solutions of Einstein equations are known, in particular, Kerr metric (see \cite{Ferrando-Saez:2009} and references therein). We need to extend these results and, especially, develop these techniques for approximate solutions. 

The above expressions, simple for a theoretical physicist, may seem complicated for an experimental one. 
They need to be broken down in measurable terms so as to conceive the devices and procedures to measure them. 
But there is no doubt that the questions that local intrinsic characterizations of metrics allow to decipher, deserve a deeper study. 

%
%
\section{Finite-differential geometry}
As it is well known, the mathematical substratum of general relativity is Lorentzian differential geometry. It thus follows that differential geometric methods are {\em unavoidable} in relativity. But there exist many situations in which these differential methods appear to be manifestly {\em insufficient}.

Among these situations, there are specific epistemic ones, involving the physical description in real time and by arbitrary observers of general variable gravitational fields, but also situations much more near of our usual ones in Earth's relativity. 

\begin{figure}[h]\label{EarthAndConstell}
\begin{center}
\includegraphics[height=3cm]{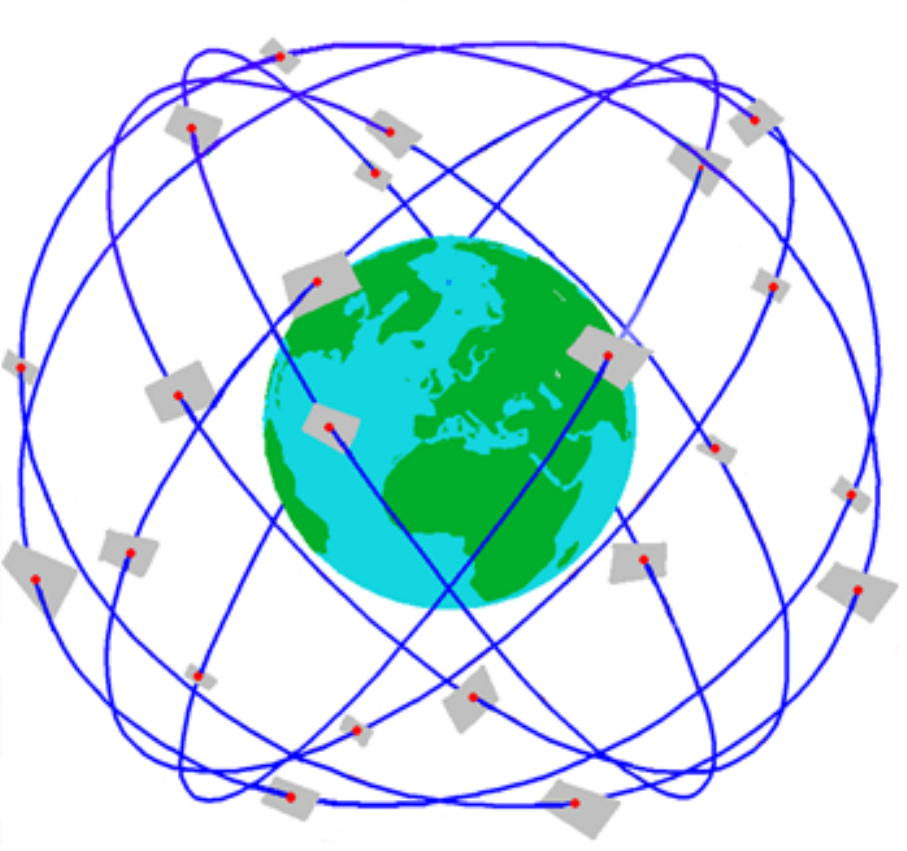}
\end{center}
\vspace{-4mm}
\caption{A constellation $\cal C$ of satellized clocks around a mass $M$, every tetrad of them working as an auto-locating system, recording the data set $\Lambda$ of all the signals emitted and received between them.}
\label{EarthAndConstell_fig:7}
\end{figure}
Thus, in a vacuum space-time, consider a mass $M$, not necessarily spherical, and a constellation $\cal C$ of satellized clocks around $M$ and such that every tetrad of them works as an auto-locating system. Let $\Lambda$
be the data set of all the signals that every clock emits and receives both, as an element of all the auto-locating systems at which it belongs and as a user of the auto-locating systems constituted by all the other clocks. 

Suppose we know the mass $M$ and the world-lines of the constellation $\cal C$. We can model the system as follows.
Start from the space-time metric around $M$, integrate the geodesic equations and specify them for every one of the clocks so as to model the whole constellation $\cal C$. Then, integrate the equations of the light-cones, either starting from the general solution of the null geodesic equations or integrating the geodesic distance function and considering its vanishing, so as to model any signal able to be emitted or received by any of the clocks of the constellation $\cal C$. With these two modeled ingredients, we are able to predict any of the data of the set $\Lambda$ received or emitted by any of the clocks in terms respectively of the data emitted or received by all the other clocks.

The physical interest of such a mathematical model is not, however, very intense. On the contrary, the following scenario could be the prelude to a fine gravimetry of the Earth ... if we were able to develop the adequate mathematical instrument. 

\begin{figure}[h]\label{ConstellC_fig}
\begin{center}
\includegraphics[height=3cm]{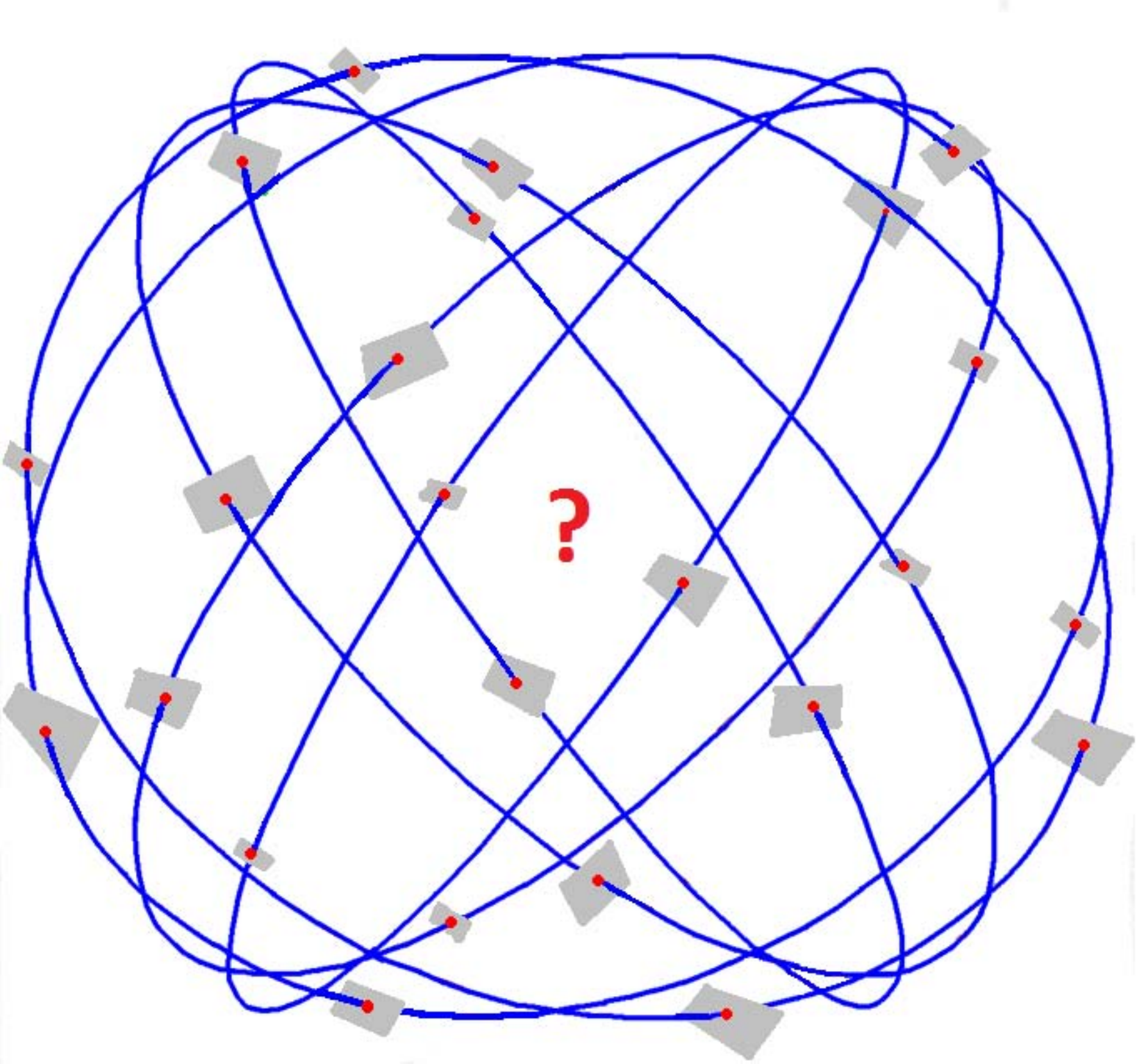}
\end{center}
\vspace{-4mm}
\caption{When we know the constellation of satellites $\cal C$ and the data set $\Lambda$, how to obtain the mass $M$?}
\label{ConstellC_fig}
\end{figure}
Suppose now that we know the constellation $\cal C$ and the data set $\Lambda$, and that we want to obtain the mass $M$. At present, the only way we know to tackle the problem is the above model, but for this scenario, it is heavy and unadapted to the starting point, the data set $\Lambda$. Indeed, $\Lambda$ consists of time-like distances (geodesic time intervals between any two events on the world-line of every clock) and null distances (light links between every pair of clocks), that is to say, of values of the {\em distance function} between pairs of events causally separated. For this reason, any {\em direct} method to model this scenario would start with the obtainment, from the data set $\Lambda$, of the metric distance function by means of a suitable interpolation method. This is part of the adequate mathematical instrument that we would be able to develop for obtaining the mass $M$. But it is still insufficient.

The objective of {\em finite-differential geometry} is to study {\em finite and interchangeable} versions of the ingredients of differential geometry (metric, connection, curvature). Because for general relativity the basic ingredient is the metric $g$, and that its finite version, the distance function, already poses problems, we shall consider it here.%
\footnote{This notion of finite-differential geometry was first presented as part of a lesson at the {\em International School on Relativistic Coordinates, Reference and Positioning Systems}, Salamanca, 2005. The mathematical results also appeared in \cite{Coll:2013}.}%

It is well known that the finite version of the metric $g$ is the {\em distance function} $D(x,y)$, or its half-square $\Omega(x,y)$, the Synge's {\em world-function}, 
\begin{equation}\label{} 
\Omega(x,y) \equiv \frac{1}{2}D(x,y)^2 \quad , 
\end{equation}
given by
\begin{equation}\label{} 
\Omega(x,y) = \frac{1}{2} \int^1_0 g\left(\frac{d\gamma}{d\lambda},\frac{d\gamma}{d\lambda}\right)d\lambda \quad , 
\end{equation}
and verifying its {\em fundamental equations}:
\begin{equation}\label{fund} 
g^{\alpha\beta}\partial_\alpha \Omega \partial_\beta\Omega = 2\Omega \quad , \quad 
g^{ab}\partial_a\Omega\partial_b\Omega = 2\Omega \quad , 
\end{equation}
where Greek indices correspond to coordinates at the point $x$ and Latin ones to the point $y$. 

Mathematically, {\em distance spaces}, i.e. manifolds endowed with a distance function, are well known,%
\footnote{The concept is due to Fr\'echet. Hausdorff named them `metric spaces' (`metrischer Raum') but in our context, it is better to call them `distance spaces'.} %
but their link with differential geometry have not been sufficiently explored. 

An important obstruction for the interchangeability between the differential and the finite concept is that the most part of distance functions {\em are not} geodesic distance functions of any metric.
It is known that, when a distance function $D(x,y)$ is the geodesic distance function of a metric, this metric may be obtained as minus the limit $y \rightarrow x$ of the mixed first derivatives of the half square (Synge's world function) of the 
distance function.%
\footnote{I am grateful to Abraham Harte for a pertinent observation on this fact.} %
If this limit is applied to an arbitrary distance function, it may give rise to a zero, degenerate or regular metric. But, even when this metric is regular, the distance function $D(x,y)$ will not be generically the geodesic distance function of it. 
Suppose, in our physical case, that the interpolated distance function from the data set $\Lambda$ is at least of differentiability class 2. We can obtain a metric from the above limiting process, but this metric will strongly depend %
\footnote{At points out of the data of $\Lambda$.} %
on the interpolation method used, for which we have no control. For this and other applications, to discern if a distance function is a geodesic distance function of a metric, it would be convenient to have an IDEAL%
\footnote{See Section \ref{Intrinsic} for this notion.} %
criterion involving solely the distance function, without any limiting process. 

I solved this problem some years ago. Let us introduce the following quantities of the first and second derivatives of the distance function $D(x,y)$:
\begin{equation}\label{Vabcalfa} 
V_{\ell mn}^\alpha \equiv \epsilon^{\alpha\lambda\mu\nu}D_{\ell\lambda} D_{m\mu} D_{n\nu}
\quad , \quad
V^{a\alpha} \equiv \epsilon^{a\ell mn} \epsilon^{\alpha\lambda\mu\nu}D_\ell D_\lambda D_{m\mu} D_{n\nu} \quad ,
\end{equation}
as well as the quantity:
\begin{equation}\label{Vabcalfa} 
V^\alpha \equiv V_{\ell mn}^\alpha x^\ell y^m z^n \; ,
\end{equation}
where $x^\ell$, $y^m$, $z^n$ are arbitrary independent directions. Define the two scalars
\begin{equation}\label{fiandpsi} 
\Phi \equiv D_\lambda V^\lambda \quad , \quad \Psi \equiv \epsilon^{r\ell mn} V_{\ell mn}^\rho D_r D_\rho 
\quad ,
\end{equation}
and form the two quantities:
\begin{equation}\label{Dsupers} 
D^\alpha \equiv \frac{V^\alpha}{\Phi} \quad , \quad
D^{a\alpha} \equiv 3 \frac{V^{a\alpha}}{\Psi} \; .
\end{equation}
Then, we have:

\vspace{3mm}
{\bf Theorem 6-1.-} (Structure theorem for geodesic distance functions.) {\em The necessary and sufficient condition for a distance function $D(x,y)$ to be the geodesic distance function of a metric, is that its derivatives verify:}
\begin{equation}\label{cnsdistancegeod}
D^\rho D_{abc\rho} + D^\rho D^{m\sigma} \oint_{abc} (D_{am\rho} - D^nD_{mn\rho}D_a)D_{bc\sigma} = 0
\end{equation}
{\em where the subscripts denote partial derivatives and $D^a$ and $D^{a\alpha}$ are the quantities just defined.}

\vspace{3mm}
In our physical case, these are the constraints to be imposed directly to any interpolated distance function on the data set $\Lambda$. Once these equations verified at the suitable degree of precision, one has to extract the metric from this distance function without any limiting process. I solved this problem together with the above one. The result is: 

\vspace{3mm}
{\bf Theorem 6-2.-} (Metric of a geodesic distance function.) {\em In terms of the partial derivatives $D_\alpha ,$ $D_{a\alpha}$ and $D_{ab\alpha}$ of a geodesic distance function, the contravariant components $g^{\alpha\beta}$ of the metric at the point $x,$ are given by}
\begin{equation}
\label{gsuper}
g^{\alpha\beta} = D^\alpha D^\beta + D^{a\alpha} D^{b\beta} D_{ab\gamma} D^\gamma \quad ,
\end{equation}
{\em where $D^\alpha$ and $D^{a\alpha}$ are the quantities above defined.}

\vspace{3mm}
Coming back to our physical case, (\ref{gsuper}) would give us the metric in the region of the constellation. Then, the intrinsic characterization method of Section 5, once generalized to perturbations of Schwarzschild gravitational field, would give us the mass $M$ that satellizes the constellation $\cal C$ and generates the data set $\Lambda$.

Theorems 6-1 and 6-2 give to the distance function $D(x,y)$ the interchangeable character with its differential homolog, the metric $g(x)$, that our finite-differential geometry requires.%
\footnote{On a metric manifold a distance function is generically {\em local} (normal geodesic domain), so that a (global) metric cannot be generically {\em interchanged} by a sole distance function, but by a suitable atlas of normal geodesic domains with distance functions submitted, in the intersection of charts, to compatibility conditions. In our case, nevertheless, the constellation $\cal C$ is contained clearly in a normal geodesic domain. Anyway, what we want here is to show the interest of the concept.} %
But finite-differential geometry has to be developed in many other directions. Among them, perhaps the more urgent ones are the development of methods of interpolation and approximation of distance functions.
\section*{Acknowledgements}
This work has been supported by the Spanish ``Ministerio de
Econom\'{\i}a y Competitividad", MINECO-FEDER project
FIS2015-64552-P.


\begin{thebibliography}{999}



\bibitem{Coll:2013} B. Coll, Relativistic positioning systems: perspectives and prospects.
{\em Acta Futura}, 7 35-47, 2013.

\bibitem{Coll-Ferrando-Morales:2010a} B. Coll J. J. Ferrando and J. A. Morales-Lladosa. Positioning systems
in Minkowski space-time: from emission to inertial coordinates. {\em Class.
Quantum Grav.}, 27:065013, 2010.

\bibitem{Coll:2001} B. Coll. Elements for a theory of relativistic coordinate systems. Formal
and physical aspects. {\em Proceedings of the Spanish Relativity Meeting 2000
on Reference Frames and Gravitomagnetism (Valladolid, Spain)}, pages
53-65, World Scientic, Singapore, 2001.

\bibitem{Coll:2006} B. Coll and J. M. Pozo. Relativistic positioning systems: the emission
coordinates. {\em Class. Quantum Grav.}, 23:7395, 2006.

\bibitem{Coll-Ferrando-Morales:2010b} B. Coll J. J. Ferrando and J. A. Morales-Lladosa. Positioning in a 
at two-dimensional space-time: the delay master equation. {\em Phys. Rev. D},
82:084038, 2010.

\bibitem{Coll-Ferrando-Morales:2012} B. Coll J. J. Ferrando and J. A. Morales-Lladosa. Positioning in
Minkowski space-times: Bifurcation problem and observational data.
{\em Phys. Rev. D}, 86:084036, 2012.

\bibitem{Coll-Ferrando:1997} B. Coll and J. J. Ferrando. The Newtonian Point Particle. {\em Proceedings
Spanish Relativity Meeting 1997 (Palma de Mallorca, Spain)}, pages 184-
190, Pub. Univ. Illes Balears, 1997.

\bibitem{Ferrando-Saez:1998} J. J. Ferrando and J. A. S\'aez. An intrinsic characterization of the
Schwarzschild metric. {\em Class. Quantum Grav.}, 15:1323, 1998.

\bibitem{Ferrando-Saez:2009} J. J. Ferrando and J. A. S\'aez. An intrinsic characterization of the Kerr
metric. {\em Class. Quantum Grav.}, 26:075013, 2009.




\end{thebibliography}
\end{document}